\documentclass[conference,letter,9pt]{IEEEtran}

\ifCLASSINFOpdf
  \usepackage[pdftex]{graphicx}     \graphicspath{{./figs/}}
      \DeclareGraphicsExtensions{.pdf,.jpeg,.png}
\else
                  \fi

\usepackage[cmex10,fleqn]{amsmath}
\usepackage[fleqn]{mathtools}
\usepackage{amssymb}
\usepackage{pifont}

\usepackage{algorithm}
\usepackage{algpseudocode}

\usepackage{array}

\usepackage[labeled,resetlabels]{multibib}
\newcites{S,P}{Supplementary References,	Postscript References}

\usepackage{float} 
\usepackage{ulem}
\usepackage{color}
\usepackage{soul}

\usepackage{fancyhdr}

\usepackage{colortbl}

\usepackage{afterpage} 
\usepackage{fixltx2e} 
\usepackage{multirow}

\usepackage{randtext}
\catcode`\_=11\relax
\newcommand\email[1]{\_email #1\q_nil}
\def\_email#1@#2\q_nil{  \href{mailto:#1@#2}{{\randomize{#1}\emailampersat \randomize{#2}}}}

\newcommand\emailampersat{{\small@}} \catcode`\_=8\relax

\usepackage{doi}

\usepackage{xfrac}

\usepackage{bm}

\makeatletter
\let\ftype@table\ftype@figure
\let\ftype@algorithm\ftype@figure
\makeatother

\newcommand{\LineComment}[1]{\State \(\triangleright\) {#1}\hfill}
\algblock{ParFor}{EndParFor}
\algnewcommand\algorithmicparfor{\textbf{parfor}}
\algnewcommand\algorithmicpardo{\textbf{do}}
\algnewcommand\algorithmicendparfor{\textbf{end\ parfor}}
\algrenewtext{ParFor}[1]{\algorithmicparfor\ #1\ \algorithmicpardo}
\algrenewtext{EndParFor}{\algorithmicendparfor}
\iftrue
\usepackage{url}
\else
\usepackage[bookmarks=false]{breakurl} \usepackage[usenames,dvipsnames]{xcolor}
\hypersetup{
    colorlinks,
    linkcolor={red!50!black},
    citecolor={blue!50!black},
    urlcolor={blue!80!black}
}
\fi

\usepackage{textcase}

\usepackage[font=small,labelfont=bf]{caption}

\usepackage[para]{footmisc}

\begingroup \makeatletter
\gdef\eqna@origamp{&} \catcode`\&\active \gdef\eqna@newamp{  \ifx\@currenvir\eqna@currenvir     \eqna@onlyfirstamp\let\eqna@onlyfirstamp\@empty   \else     \eqna@origamp   \fi
}
\gdef\eqna@hook{  \let\eqna@currenvir\@currenvir   \catcode`\&\active   \let&\eqna@newamp   \let\eqna@onlyfirstamp\eqna@origamp   }
\catcode`\*11 \gdef\eqnarray{\eqna@hook\align} \gdef\eqnarray*{\eqna@hook\align*} \global\let\endeqnarray\endalign
\global\let\endeqnarray*\endalign*
\endgroup

\usepackage{tikz}
\usetikzlibrary{calc,fit,arrows,shapes.misc}

\usepackage{spreadtab}
\STautoround*{2} 
\usepackage{numprint}
\npthousandsep{,}\npthousandthpartsep{}\npdecimalsign{.}

\def\por1{\partial}

\usepackage[binary-units=true,
prefixes-as-symbols=false,
]{siunitx}
\DeclareSIUnit{\microsecond}{\SIUnitSymbolMicro s} 
\newcolumntype{H}{>{\setbox0=\hbox\bgroup}c<{\egroup}@{}}
\newcolumntype{M}{>{\centering\arraybackslash}m{\dimexpr0.25\linewidth-2\tabcolsep}} \newcolumntype{N}{>{\centering\arraybackslash}m{\dimexpr0.10\linewidth-2\tabcolsep}}
\newcolumntype{Y}{>{\raggedleft\arraybackslash}X}

\usepackage{caption}
\DeclareCaptionFormat{myformat}{#1#2#3\hrulefill}
\newcommand\litem[1]{\item{\bfseries #1:}}

\begin{document}
\bstctlcite{IEEEexample:BSTcontrol} 
\title{FairGA: Fair Genetic Algorithm ---\\ Beyond Resource-oriented Sustainability for ICT Products and Services}  
 \author{\IEEEauthorblockN{\small Reza \MakeTextUppercase{Farrahi Moghaddam}$^{1,2,*}$}
\IEEEauthorblockA{$^{1}$Synchromedia Lab and CIRROD}
\IEEEauthorblockA{ETS (University of Quebec)}
\IEEEauthorblockA{Montreal, QC, Canada H3C 1K3} 
\IEEEauthorblockA{Email: \email{imriss@ieee.org}} \IEEEauthorblockA{LinkedIn: \url{https://www.linkedin.com/in/rezafm}}
\IEEEauthorblockA{$^{*}$Corresponding author} \and
\IEEEauthorblockN{\small  Yves \MakeTextUppercase{Lemieux}$^{2}$}
\IEEEauthorblockA{$^{2}$Ericsson Research - Cloud Technology}
\IEEEauthorblockA{Ericsson Canada Inc}
\IEEEauthorblockA{Montreal, QC, Canada H4P 2N2} \and
\IEEEauthorblockN{\small  Mohamed \MakeTextUppercase{Cheriet}$^{3}$}
\IEEEauthorblockA{$^{3}$Synchromedia Lab and CIRROD}
\IEEEauthorblockA{Prodcution Automation Department}
\IEEEauthorblockA{ETS (University of Quebec)}
\IEEEauthorblockA{Montreal, QC, Canada H3C 1K3} 
}

\maketitle

\begin{abstract}
The complexity of ICT products and services has brought them to the level of disposable `species'. The combination of the race to optimal performance and disposability has resulted in considerable footprint and impact. Although approaches such as increasing efficiency, reducing the total cost of ownership, life cycle assessment and management, and circular economy have been put forward to manage and reduce the footprint and impact, the complexity of processes involved and especially invisibility of key but unobservable processes has resulted in some lower bounds for minimal achievable footprint. In this work, a modified approach to the Genetic Algorithm is proposed in order to introduce the notion of `nondisposability' to the ICT products and services in order to implicitly influence and manage unobservable processes, and ultimately reduce the overall footprint and resource consumption. The proposed genetic algorithm is called FairGA, and it is compared with the traditional Genetic Algorithm against standard optimization functions. Also, the impact of the FairGA on the resource extraction has been illustrated with promising results.
\end{abstract}

\IEEEpeerreviewmaketitle

\section{Introduction}
\label{sec_Introduction}
In the context of sustainability, the Information and Communications Technology (ICT)\footnote{In order to explicitly include the embedded cases, we would prefer to call it (E)ICT: The (Embedded) Information and Communications Technology \cite{Farrahi2014c}.} has been recognized as a key for a sustainable future by providing functions (services) with minimal to zero physical impact \cite{Malmodin2015,SMARTer2030}. However, the hidden `footprint' of the ICT itself may cancel out its benefits, and therefore there has been a great effort in order to contain such footprint \cite{Simeonidou2011,Vereecken2010,Farrahi2014f}. Many approaches and tools, such as Life Cycle Assessment (LCA) \cite{Arushanyan2014,Honee2012,Farrahi2015b} and Circular Economy \cite{Turner1992,Andersen2007},\footnote{In addition to other (indirect) approaches such as Efficiency and Total Cost of Ownership (TCO) \cite{Mainstay2016}.} have been considered in order either to identify all footprint and then reduce it or to displace the footprint back into the new products and services. 

The complexity of ICT products and services is continuously increasing which should be also combined with the complexity of their context (market and the ecosystem in general). This has been resulted in appearance and rapid disposal of various products (services) from the markets. These waves have been usually driven by interest, profit, or `disruption', and they do not factor in the associated footprint/impact related to `short-lived' disposable products (services). Although measures such as recycling have been of great interest to compensate the impact, they usually expensive, hard to verify, and of low performance. 

In this work, we propose an optimization approach that helps to influence the unmeasurable/unobservable processes\footnote{including but not limited to hidden or latent processes.} in an operation. Considering the fact that all actors involved in the life cycle of a product/service would seek optimization of their goals and indicators, application of the proposed optimization approach would help to reduce the `total' footprint and impact associated to their operations even if there are processes invisible to them. We used the Genetic Algorithm (GA) \cite{Mitchell1998} as the baseline of our approach considering its performance in the context of ICT-related problems \cite{Farrahi2014b}. However, the same vision could be applied to any other optimizer. The core of our proposed optimization, called the Fair GA (FairGA) optimization, is introducing a minimal life time for every product/service (a chromosome in the context of the optimization process). The introduced constraint forces the `hidden' processes to implicitly react and lower the volume of production manufactured.

The paper is organized as follows. In Section \ref{sec_Genetic_Algorithm_Pure_Optimization}, the overview of a generic GA optimizer is provided. Then, the challenges namely non-resource processes and invisible processes are discussed in Sections \ref{sec_The_Challenge_of_Non_Resource_Entities} and \ref{sec_The_Challenge_of_Dark_Matter}. The proposed FairGA approach and one of its algorithm instances is presented in Section \ref{sec_Fair_GA_FairGA_Optimization}, followed by the experimental results in Section \ref{sec_Experimental_Results}. The positive impact of the proposed FairGA in improving operations and in particular reducing the resource extraction volume is discussed in Section \ref{sec_Discussions_FairGA_Potential_Role_in_Sustainability}. Finally, the conclusions and the future prospects are presented in Section \ref{sec_conclusions}.

\section{Genetic Algorithm: Pure Optimization}
\label{sec_Genetic_Algorithm_Pure_Optimization}
The Genetic Algorithm (GA) is well-known as a metaheuristic optimization approach that is based on `natural' processes namely crossover and mutation in order to search for a global optimal \cite{Mitchell1998,Chu1997,Farrahi2014b}. Although there are various GA approaches, a common characteristic of all them is that `a' chromosome per se does not carry any `value' other than its potential contribution to the search for the global optimum. In other words, a chromosome in any of the GA approaches is `disposable'. This is the main feature that differentiate between the proposed FairGA vision compared to the other GA approaches, and in the proposed approach a chromosome is not available for immediate disposal. We discuss this feature in details in the next subsection.

\subsection{Optimization with respect to ``Objectives''}
\label{sec_Optimization_with_respect_to_Objectives}
As mentioned before, `objectives' are the core and essential part of any optimization, including that of the GA optimization. However, the actions and operations taken toward reaching the global optimum with respect to the objectives require some resources and in general bear some costs. Usually, the cost of operation toward getting to the optimal state\footnote{In some cases, the state of a system/ecosystem could be represented by a `point' in a space, which is then called the state space.} with respect to the objectives is calculated in a forward-looking way, i.e., any operation that does no contribution (or negative contribution) to the journey toward the optimal state would be simply discarded and avoided. One of common actions that is avoided is keeping the low-value chromosomes in the population because they contribute little to the optimization and at the same time add additional costs such as calculation of the objective function. 

However, every chromosome carries other values more than that of their contribution to the optimization. These values become more relevant when the non-resource entities,\footnote{An entity is a thing that can have distinguishable boundaries with respect to other things and entities. For example, `autonomous' things are by nature entities.} i.e., those entities that cannot be equivalently decomposed into only resource subparts, are the actual physical things that the chromosomes represent. In the next section, the importance of inclusion of this form of chromosomes and value in the optimization is discussed.

\section{The Challenge of Non-Resource Entities}
\label{sec_The_Challenge_of_Non_Resource_Entities}
Not all operations involve ``resource'': Although it has been a trend to account for all elements used in the processes of an operation in the form of resource, there are more and more instances where this approach would not be applicable. For example, when an entity could be decomposed into its substances such as the chemical elements, does it mean that that entities should be considered as a resource of those chemical elements? This may be valid in those cases where the ``composition'' itself does not carry any ``value''. However, it seems that this will not be a general case in the future, and the negligence of the {\em value} would be an unfair approach. We leave further discussion on this topic to a future work, and only summarize it in the form of an assumption in this work: 

{\em Assumption 1: ``Every entity has a value and therefore it is entitled to a fair handling.''}

Before providing the definition and algorithm of the FairGA, another important aspect of control and management of the system/ecosystem is discussed in the next section with a focus on unobservable processes and areas of operation.

\section{The Challenge of `Dark Matter'}
\label{sec_The_Challenge_of_Dark_Matter}
It is a well-known principle that `You cannot manage what you do not measure' \cite{Fitzenz2002}. Although this is in general correct, there is a possibility of risk in its practicing when the boundaries of the system are not known or are not set properly. Even if the `boundaries' are well set, if some regions are not measurable (observable), then the whole optimization would enter a state of stall because there would be no differentiation between various possible operation (trajectories) starting from the current state with respect to optimization of objectives.  

Even in cases where the objectives are 'measurable', there is still a possibility of risk especially for those systems/ecosystems that possess ``inertia'', i.e., they cannot rapidly displace from a particular state to a desired state. A worst case scenario in this context would be observation of `incremental' increase in the objective along a trajectory of changes in the state while the states themselves approach a highly non-optimal state. This would end in a sharp drop in the optimization objective when the system/ecosystem reaches the non-optimal state. There would be then associated costs to bring the system back on an optimal trajectory.

To represent these uncertainties, we introduce the notion of {\em dark matter}, which accounts for all those regions of the system/ecosystem within its boundaries that are not ``observable''. We use the term observable in a generalized form that covers both hidden and also visible but unmeasurable regions. Also, the condition of having dark matter within the boundaries is actually a definition for the boundary. Therefore, if there is a dark matter outside the current assigned boundary of a system/ecosystem, the boundary should be adjusted to include that region instead of discarding it.

The main feature of dark matter is that it is unmeasurable but influential. Although modeling the dark matter would be impossible, we argue that if we `introduce' a form of {\em artificial} ``physics'' (which imposes its limitations) in the rest of the system/ecosystem, the dark matter would adapt/adjust to that induced physics, and therefore there would be means to `influence' and manage the dark matter and then to  achieve a greater outcomes/behavior of the whole system/ecosystem without actually knowing the {\em true} physics/model of the dark matter. 

The Fair GA (FairGA) optimization that will be introduced in the following sections could be seen as a primitive way to inject such limitations. Although we will postpone studying the application of the FairGA in the context of management of unobservable regions, it is worth mentioning that the application of the FairGA for such a purpose would be probably in the form of calculating a from of the Total Cost of Action (TCoA) instead of the Execution Cost of Action (ECoA). The ECoA is all the costs required to be paid in order to perform/execute an action, while the TCoA also includes other associated costs (such as post-action costs) of the action in addition to the ECoA. Also, the term cost itself is used in a generic way that includes all forms of {\em providing}, including and beyond money. We will explore this aspect in greater details in another work.

\section{Fair GA (FairGA) Optimization}
\label{sec_Fair_GA_FairGA_Optimization}
In this section, a form of the Fair GA (FairGA) optimization is introduced. Although this form could be taken as the definition of the FairGA, we let the possibility for more generic definition open for the future. For now, we  {\em weakly} define FairGA optimization as an optimization that `fairly' handles the participants (chromosomes) in the GA populations.

To introduce the proposed form of the FairGA, we start with a few assumptions and definitions. Also, from here on, we refer to the proposed form as FairGA itself for the purpose of simplicity. 

\subsection{FairGA: Assumptions and Definitions}
\label{sec_FairGA_Assumptions_and_Definitions}
In addition to the definitions provided in Section \ref{sec_Genetic_Algorithm_Pure_Optimization}, a few additional definitions related to the FairGA are provided below:
\begin{enumerate}
	\litem{Chromosome Life Time ($L_i$)} For each individual chromosome $i$ in the population, the number of iterations that chromosome participates (stays) in the population is defined as its Life Time. Considering the complexity of the systems/ecosystems under optimization, there is a highly small chance that a specific chromosome appears more than once in the population. However, in the event of reappearance, the second entry is not added up to the first one, and it is considered as {\em another} individual chromosome itself.
	\litem{Population Size ($S$)} At each iteration (time instance), the total number of participating chromosomes in the population $P$ is defined as the Population Size $S$ of that instance/iteration.
	\litem{Max Population Size ($S_\text{max}$)} The maximum allowed Population Size at any time is defined as Max Population Size. Compared to the GA optimizations, the Max Population Size of FairGA is equal to their actual population size.
\end{enumerate}
There is also a few more assumptions:
\begin{enumerate}
	\litem{Min Chromosome Life Time ($L_\text{min}$)} (Assumption 2) In the proposed form of the FairGA in this work, every chromosome that enters the population is `assumed' to be {\em fairly} handled if its Life Time is {\em greater or equal to} a preset parameter defined as Min Chromosome Life Time: $L_i \geq L_\text{min}, \forall i\in P$.
	\litem{Disposable Population Size ($S_\text{dispose}$)} (Assumption 3) At any iteration, a group of chromosomes of a maximum size of Disposable Population Size could be discarded. It is considered that the discarded chromosomes have been fairly handled before the event of discard.
	\litem{Population Refresh Ratio ($\rho$)} The ratio of Disposable Population Size to Max Population Size is defined Population Refresh Ratio: $\rho=S_\text{dispose} / S_\text{max}$. 
\end{enumerate}

\subsection{FairGA: The Algorithm}
\label{sec_FairGA_Algorithm}
The FairGA approach could be implemented in various forms and algorithms. The particular FairGA optimization algorithm presented in this work consists of three stages: i) The ramp-up stage, ii) The core stage, and iii) The exit stage. Each stage are separately discussed in the following subsections (and summarized in Algorithm \ref{alg_FairGA_1}). 

\subsubsection{FairGA: The Ramp-up Stage}
\label{sec_The_Ramp_up_Stage}
In contrast the GA, in the FairGA, the population cannot start at the size of Max Population Size. Otherwise, the population would be needed to be frozen for a duration of Min Chromosome Life Time, after which some chromosomes could be discarded to have new chromosomes added to the population. In order to avoid population freeze, the FairGA considers a ramp-up stage in which the rate of additions to the building up population is constant.\footnote{Nonlinear ways to execute the ramp-up stage could be also considered. However, in this work, we assume a linear ramp-up.} The rate, denoted $R_\text{Rampup}$, is assumed to be less than Population Refresh Ratio, i.e., the maximum chromosomes that can be added at an iteration should be less than Disposable Population Size. This condition actually shows the fourth (but not essential) assumption used:

{\em Assumption 4: ``Neither the population nor part of it should become frozen in an iteration.''}

It is worth mentioning that Assumptions 1 to 4 actually forming the `artificial' physics referred to in Section \ref{sec_The_Challenge_of_Dark_Matter}. When the presumed mutation/crossover process proceeds to create a new chromosome to be added to the population, the process knows that that chromosome would not exit the population at least for a number of Min Chromosome Life Time $L_\text{min}$ iterations. We will discuss more the impact of the artificial physics on the unobservable parts of the system/ecosystem in a future work.

At the end of the ramp-up stage, the population reaches $S_\text{max}$. From that instance/iteration, the second stage, the core stage, starts.

\subsubsection{FairGA: The Core Stage}
\label{sec_The_Core_Stage}
In the core stage, the population size is kept constant and equal to $S_\text{max}$. In each iteration, a group of a maximum size $S_\text{dispose}$ (equal to a portion of $\rho$ of the population) is discarded, and is replaced by the newly-generated (or existing) chromosomes that mutation/crossover process provides. The discarded chromosomes should have been fairly handled, i.e., have spent at least $L_\text{min}$ iterations in the population. The discarding criteria is based on the performance against the objective(s).

\subsubsection{FairGA: The Exit Stage}
\label{sec_The_Exit_Stage}
The exit stage is simply a waiting stage in it those chromosomes not have reached to their $L_\text{min}$ are kept alive. 
It is not expected that the optimum point would change in this stage.

\begin{algorithm}[!htbp]
		\caption{A FairGA algorithm for optimization.}
	\label{alg_FairGA_1}
	\begin{algorithmic}[1]
		\Procedure{$\text{Optimal Population} = \mathbf{FairGA}$}{Objective Function, Seed Function, Selection Function, Crossover Function, Mutation Function, $S_\text{max}$, $L_\text{min}$, $S_\text{dispose}$, $R_\text{Rampup}$, $N_\text{max}$}
		\LineComment{\emph{Initialize the population at $t=0$ ($t$ is the iterations counter).}}
		\State $P \gets$ The union of the outputs of the Seed Function called $R_\text{Rampup} S_\text{max}$ times;
		\State $F \gets$ The outputs of the Objective Function on $P$; $F$ has the same `order' as $P$;
		\State $A \gets$ A vector of 0 values of the same size and order as $P$;
		\LineComment{\emph{The Ramp-up stage.}}
		\Repeat
			\LineComment{\emph{Aging.}}
			\State $A \gets A + 1$ 
			\LineComment{\emph{Add $R_\text{Rampup} S_\text{max}$ more chromosomes to the population.}}
			\State $P \gets$ $P \Cup$ the output of $R_\text{Rampup} S_\text{max}$ calls to the Seed Function;
			\State $F \gets$ $F \Cup$ The outputs of the Objective Function on the new members of $P$;			
			\State $A \gets$ $A \Cup$ a vector of 0 values of the same size of the new members of $P$.
			\LineComment{\emph{Perform GA Operations.}}
			\State $P \gets$ The results of Selection, Crossover, and Mutation Functions with condition $L_i \geq L_\text{min}$;
		\Until{$S==S_\text{max}$.}
		\LineComment{\emph{The Core stage.}}
		\Repeat
			\LineComment{\emph{Aging.}}
			\State $A \gets A + 1$ 
			\LineComment{\emph{Perform GA Operations.}}
			\State $P \gets$ The results of Selection, Crossover, and Mutation Functions with condition $L_i \geq L_\text{min}$;
		\Until{$t==N_\text{max}-L_\text{min}$.}
		\LineComment{\emph{The Exit stage.}}
		\Repeat
			\LineComment{\emph{Aging.}}
			\State $A \gets A + 1$ 
		\Until{$t==N_\text{max}$.}				
		\State {\bf Return} $P$ and $F$
		\EndProcedure
	\end{algorithmic}
\end{algorithm}

\section{Experimental Results}
\label{sec_Experimental_Results}
Two illustrative functions have been used to compare the performance of the FairGA algorithm and a traditional GA algorithm:
\begin{enumerate}
	\litem{Berlich Noisy Parabola function} This function is defined as follows:\\ $f_1(x_1, x_2)=\left(\cos \left(\sum_{i=1}^{2}x_i^2\right) + 2\right) \left(\sum_{i=1}^{2}x_i^2\right).$\\
	This function has a global minimum point at $(0, 0)$.
	\litem{Schwefel function} This function is defined as follows: $f_2(x_1, x_2)=-1/2\left(\sum_{i=1}^{2}\left(x_i \sin \left(\sqrt{|x_i|}\right)\right)\right).$\\
	This function has a global minimum point at $(420.969, 420.969)$.
\end{enumerate}

\begin{figure*}[tbh!]
	\centering
	\begin{tabular}{@{}cc}
		\fbox{\includegraphics[height=3in]{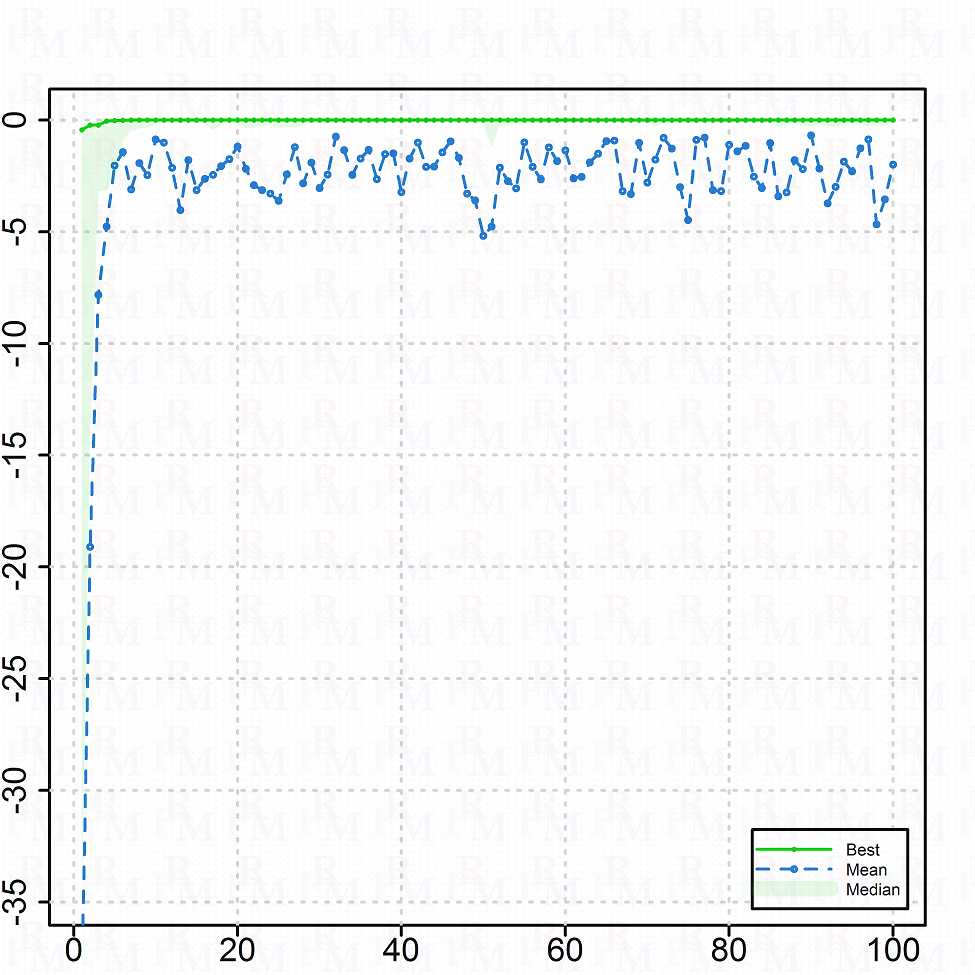}} &
		\fbox{\includegraphics[height=3in]{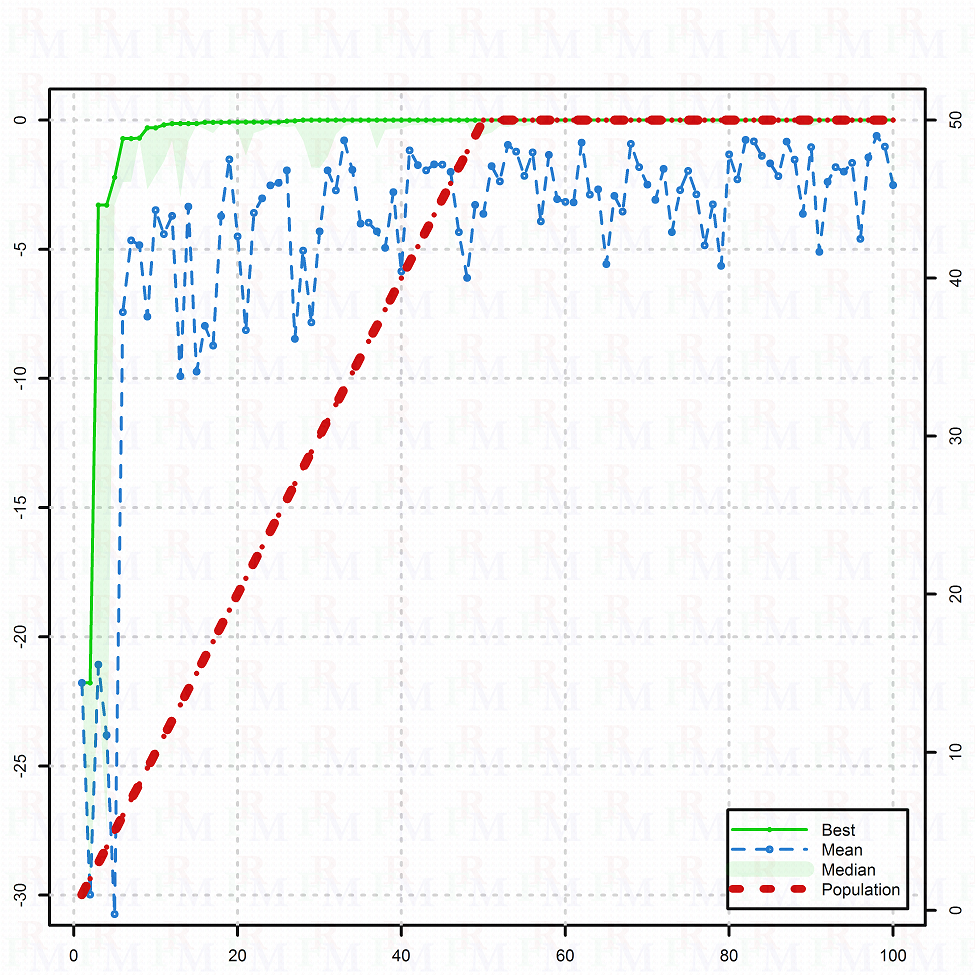}} \\
		(a) & (b) 
	\end{tabular}
	\caption{Performance comparison of FairGA and GA for the Berlich Noisy Parabola function $f_1(\cdot)$.
		a) The performance of FairGA.
		b) The performance of GA.}
	\label{fig_FairGA_GA_Comparison1}
\end{figure*}

\begin{figure*}[tbh!]
	\centering
	\begin{tabular}{@{}cc}
		\fbox{\includegraphics[height=3in]{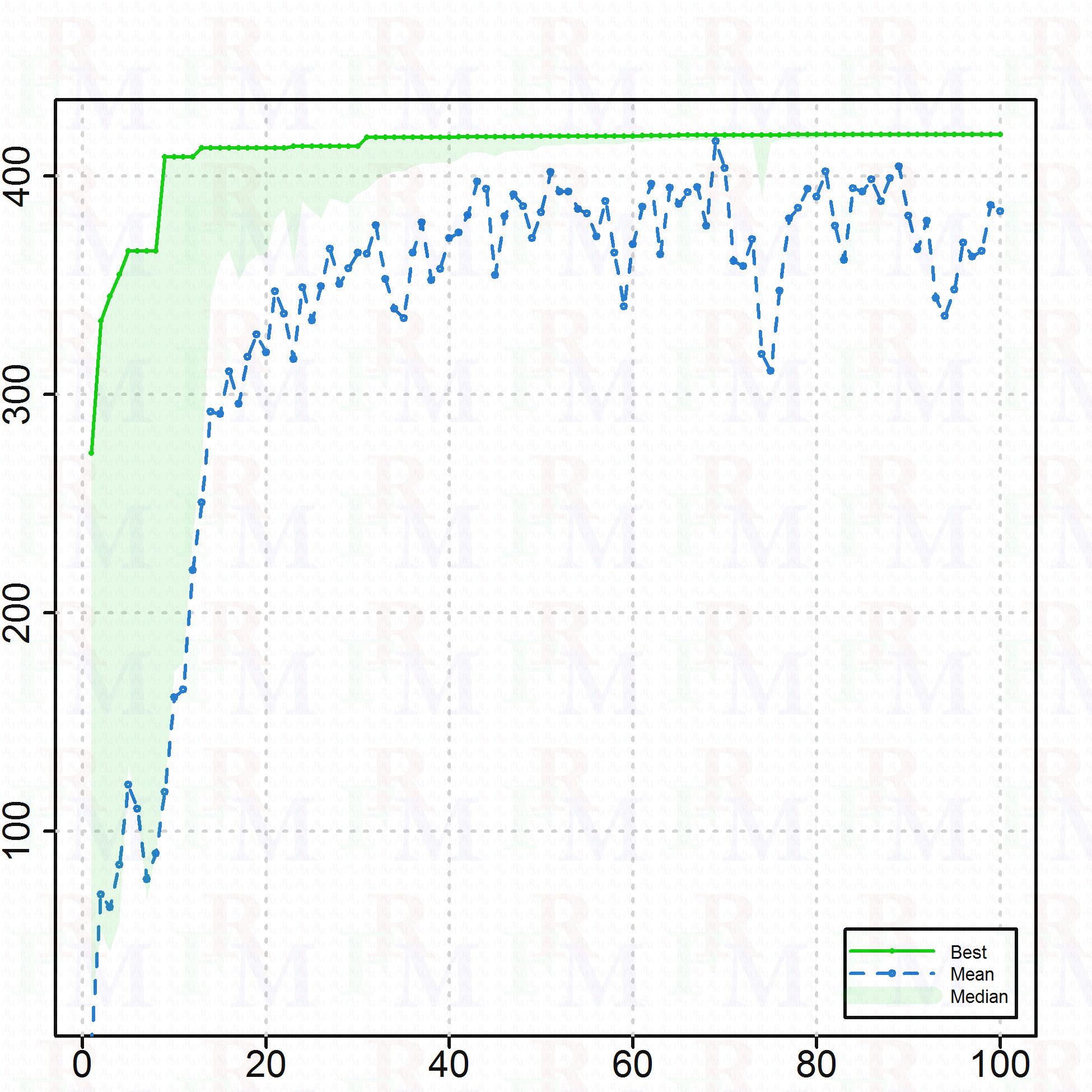}} &
		\fbox{\includegraphics[height=3in]{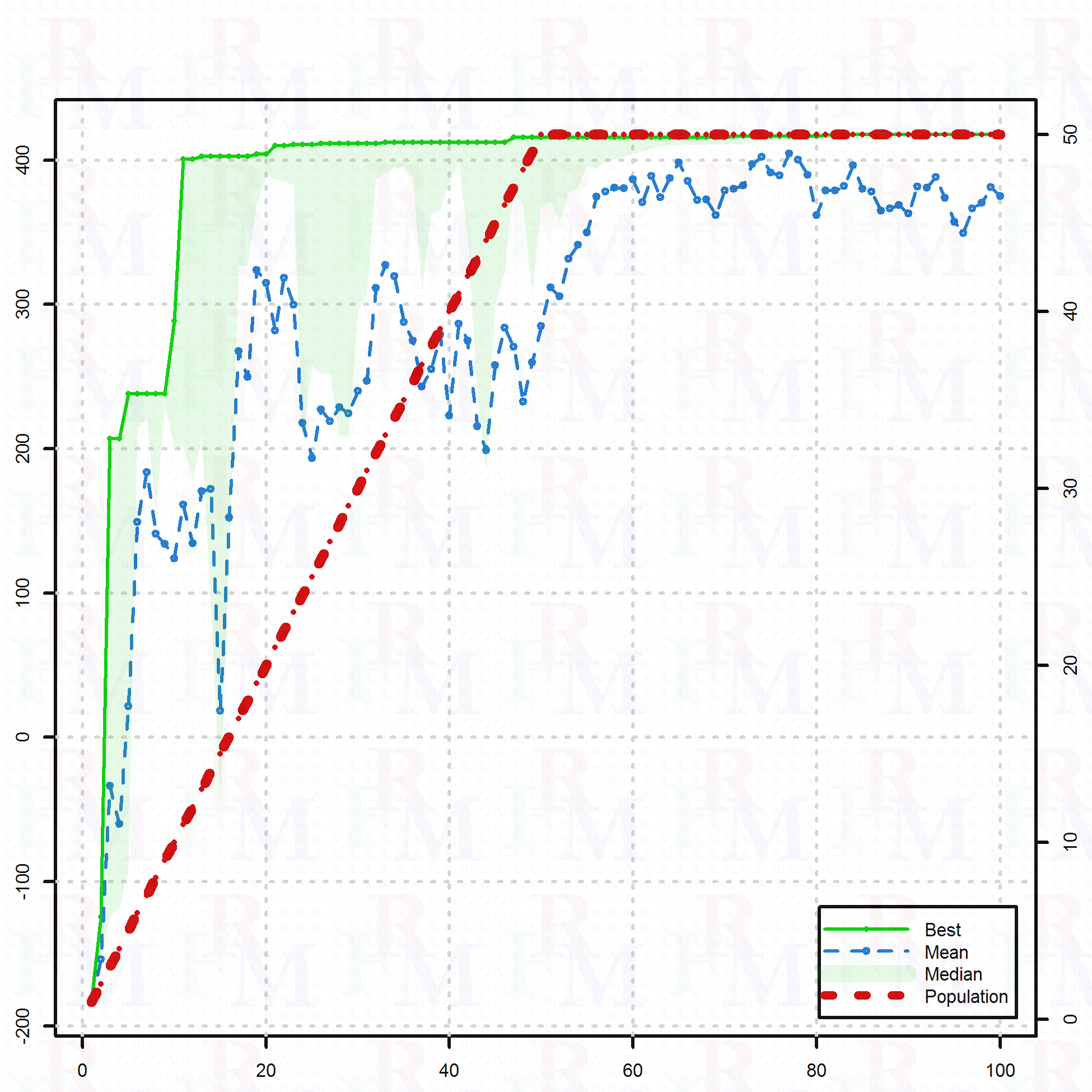}} \\
		(a) & (b) 
	\end{tabular}
	\caption{Performance comparison of FairGA and GA for the Schwefel function $f_2(\cdot)$.
		a) The performance of FairGA.
		b) The performance of GA.}
	\label{fig_FairGA_GA_Comparison2}
\end{figure*}

Figures \ref{fig_FairGA_GA_Comparison1} and \ref{fig_FairGA_GA_Comparison2} show the comparative performance of the algorithms for these two functions, respectively. A crossover rate of 80\% and a mutation rate of 10\% is considered in all cases.
Note that the actual `time' for the completion of optimization is different from the total number of iterations, especially for the FairGA. To be specific, in the case of examples used in this section, with a maximum number of iteration of $100$ and a maximum population size of $50$, the linear completion time (to the maximum number of iterations) for the GA is $100*50=$5,000 a.u. (arbitrary units) of objective function calculation, while it is $100/2*50=$3,750 a.u. for the case of the FairGA. Also, for the criteria of reaching the optimal value and considering Figure \ref{fig_FairGA_GA_Comparison2}, the linear time for the GA after 30 iterations is $30*50=$1,500 a.u., while for the FairGA it is $46/2*50=$1,150 a.u. after 46 iterations. This performance may not always be the same, and Figure \ref{fig_FairGA_GA_Comparison1} shows an example where the GA reaches the optimum in 4 iterations (equal to $4*50=$200 a.u.) much less than that of FairGA ($15/2*50=$375 a.u.). It is worth mentioning that the goal of the FairGA is not to perform better compared to the GA in terms of optimization. Instead the goal is to add fairness to the chromosomes. Nonetheless, contributions to the optimization performance is an added value for the FairGA.

\section{Discussions: FairGA's Potential Role in Sustainability}
\label{sec_Discussions_FairGA_Potential_Role_in_Sustainability}
Among various tools and approaches to total assessment and management of sustainability performance of an operation, Life Cycle Assessment (LCA) and Circular Economy have been of great interest. The LCA approaches enable the examiner to include all stages of the life cycle of a product (or a service) in the assessment in order to identify those stages or processes that produce the most footprint or consume the most resources. However, there are challenges for the LCA approaches: 
\begin{enumerate}
	\litem{Relativity} The LCA approaches by definition are relative \cite{Henriksson2014,Guinee2016,Farrahi2014c}. They allow identification and then probably management of non-optimal stages and processes. However, this does not guarantee that the `total' footprint would eventually reach the `zero' level. In other words, they could reach a `local' optimum point of operation without having any indication that it is a local optimization. Although very detailed LCA analyses would reduce the probability of falling in a local minimum, the complexity and uncertainty of operation and especially their ecosystemic nature would lead the analysis to stop at a `boundary' in order to make the calculations feasible. Moreover, the LCA analyses performed on different settings would result in `incompatible' results that would be of no use in order to rank the settings against each other, a step that is required to identify the most sustainability-oriented setting.
	\litem{Uncertainty} In addition to relative nature of the LCA analyses, their data-driven nature usually requires accounting for the  uncertainty in the calculations \cite{Beltran2016}. In some cases, this would end up in non-conclusive results that would make the LCA analyses unable to promote a sustainable operation against the others while such differentiation is of great importance. \end{enumerate}

In terms of the Circular Economy, there are also some challenges:.
\begin{enumerate}
	\litem{Invisible Stages} Although circular behavior could be directly validated from the `topology' of the operation, the inputs and outputs to intermediate stages, especially those which are not well visible, could cancel out the benefits of the circular operation \cite{Planing2015}.
	\litem{Low Circular `Flow'} The key parameter in the circular operation in the minimum flow volume across the circuit \cite{Pin2007,Ghisellini2016}. In other words, the actual flow returned to the point of start is more important than the capability to close the path to make a circuit.
\end{enumerate}

\begin{figure*}[tbh!]
	\centering
	\begin{tabular}{@{}cc}
		\fbox{\includegraphics[height=2.2in]{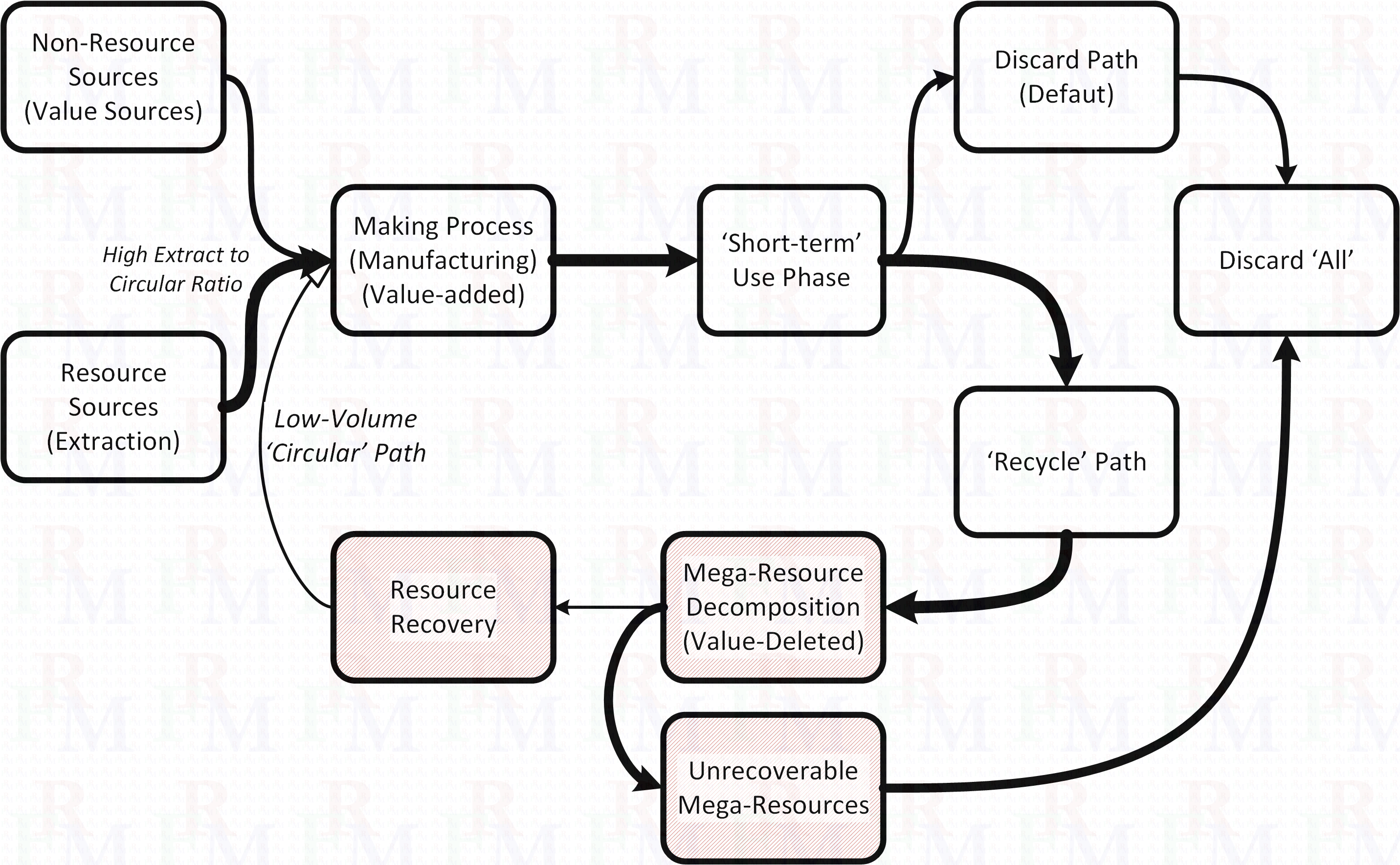}} &
		\fbox{\includegraphics[height=2.2in]{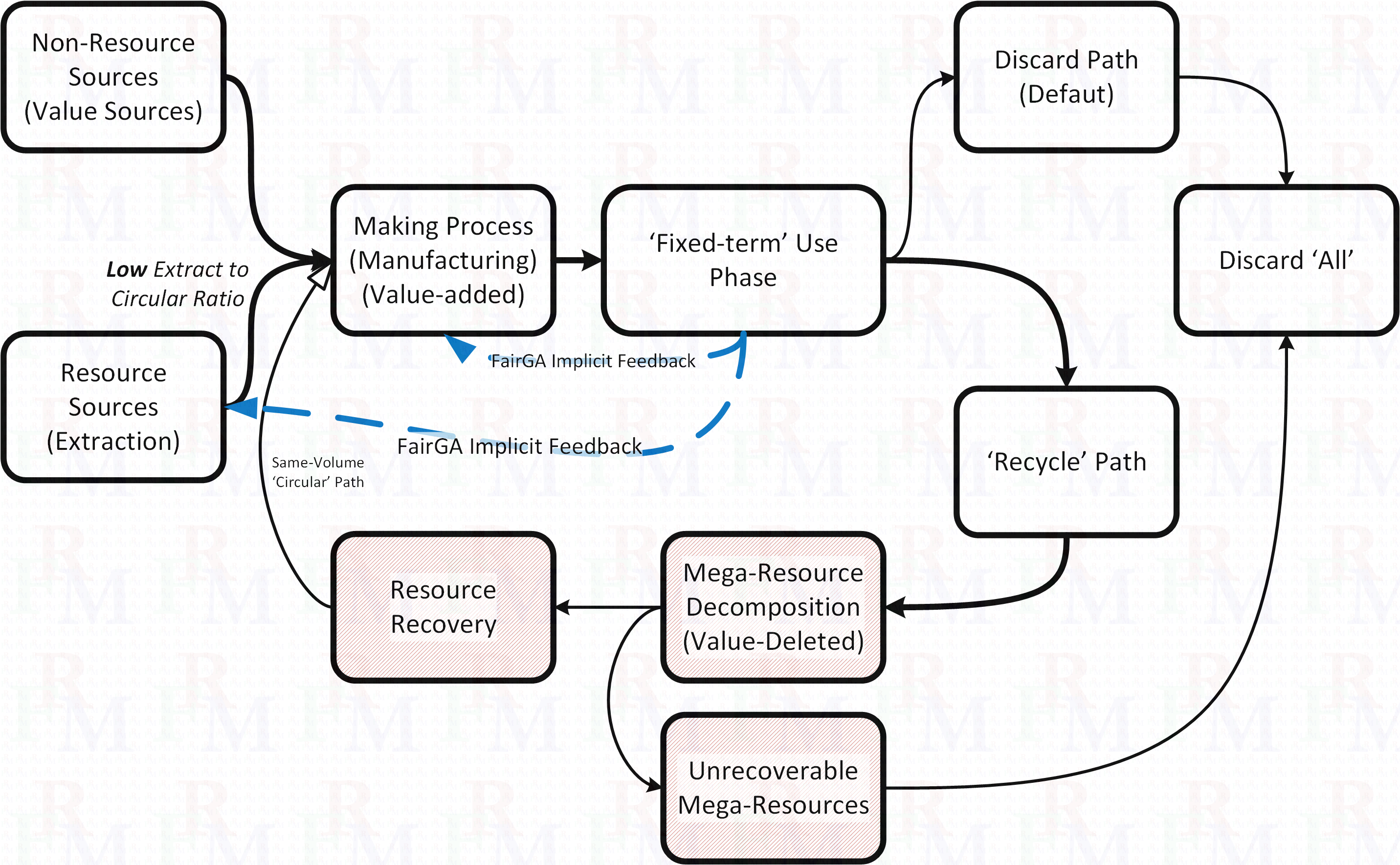}} \\
		(a) & (b) 
	\end{tabular}
	\caption{The impact of FairGA-based operation on sustainability.
		a) The circular performance with GA.
		b) The circular performance with FairGA.}
	\label{fig_FairGA_Sustainability1}
\end{figure*}

To have a simplified estimation of the impact of the FairGA or similar approaches in the lowering the overall resource extraction,\footnote{not mentioning the other benefits in terms of retaining and adding `values'.} a rate-based model is used in Figure \ref{fig_FairGA_Sustainability1}. Let us assume that resource rate (the amount of resource used to build one unit of product/service) is $\text{Res}_\text{r}$, the volume of products/services in use is $\text{Prod}_{vol}$, the rate of recycling is $\text{Recy}_\text{r}$, the rate of resource recovery from recycled products is $\text{Res}_{rec}$. Also, let us assume that in the baseline scenario, i.e., without the FairGA enforcement, the average life-time of a product in operation is $\text{Life}_{0}$ in an arbitrary unit of time (a.u.). This means that in a quasi steady-state, the volume resources returned to the manufacturing stage per one a.u. of time, which is the actual benefit from the circular economy, would be $\text{Res}_\text{r} \text{Prod}_{vol} \text{Recy}_\text{r} \text{Res}_{rec} / \text{Life}_{0}$. That means that the volume of resources to be extracted and fed into the manufacturing stage would be $\text{Res}_\text{r} \text{Prod}_{vol} \left(1 - \text{Recy}_\text{r} \text{Res}_{rec}\right) / \text{Life}_{0}$. Now, if we add the FairGA approach in the model in the form of requiring a minimum life time of $\text{Life}_{FairGA}$ for every product, the actual extracted resource volume per one a.u. of time would be $\text{Res}_\text{r} \text{Prod}_{vol} \left(1 - \text{Recy}_\text{r} \text{Res}_{rec}\right) / \text{Life}_{FairGA}$. Comparing the two scenarios, the resource extraction volume would be reduced by a factor equal to $\text{Life}_r = \text{Life}_\text{FairGA} / \text{Life}_{0}$. To have an estimation of this impact, let us assume $\text{Res}_{rec}=0.8$, $\text{Recy}_\text{r}=0.3$, and $\text{Life}_r=1.5$, which means 80\% of products are recycled, 30\% of recycled products could be recovered into resources, and the extended life time enforced by the FairGA is 50\%. In this situation, the raw-material resource extraction would be ${\small \text{Res}_\text{r} \text{Prod}_{vol} \left(1 - 0.8 * 0.3\right) / \text{Life}_{0}=0.76 \text{Res}_\text{r} \text{Prod}_{vol} / \text{Life}_{0}}$ and ${\small \text{Res}_\text{r} \text{Prod}_{vol} \left(1 - 0.8 * 0.3\right) / \text{Life}_{0} / 1.5=0.51 \text{Res}_\text{r} \text{Prod}_{vol} / \text{Life}_{0}}$, respectively for no-FairGA and FairGA scenarios. It is worth mentioning that the volume of returned resources back to the manufacturing stage by the circular economy nature of the operation is also different for the two scenarios:  $0.24 \text{Res}_\text{r} \text{Prod}_{vol} / \text{Life}_{0}$ and $0.16 \text{Res}_\text{r} \text{Prod}_{vol} / \text{Life}_{0}$, respectively. This means that the associated costs related to the processing of the recycled material is also much lower in the case of FairGA. 
The advantage of the FairGA approach comes from the fact that the constraint on the number of products that could enter the operation propagates back to the `invisible' areas of the operation, such as research and design, because `everybody' involved would like to reach a comparable performance compared to that of the baseline case (the case without any constraint on the life time of products). In other words, instead of a cruel and primitive section process, understanding of how the structure of a chromosome impacts its performance is required in the form of scientific and technological modeling and advancements. The benefits of such understanding would be much bigger than that mentioned here, especially in design of products/services (chromosomes) with unforeseen functionalities and purposes. it is also worth mentioning that the value/output of a product/service usually comes from other products/services/apps that stand on top of that product/service. Therefore, the FairGA approach also implicitly forces those out of the model services/apps to optimize themselves (the software/firmware) in order to increase the performance while the underlaying product/service does not change.

Although forcing some products stay in operation while other products with improved performance could be brought in could be seen to be in contradiction with the goal of lower footprint and impact. However, it would actually serve as an important drive of change toward better alternatives, such as modular-by-design products or disaggregated hardware \cite{Farrahi2016b,Mainstay2016}. We leave such a discussion for a future work.

\section{Conclusion}
\label{sec_conclusions}
A modified approach to genetic algorithm (GA) approaches to optimization has been proposed. The proposed approach, called Fair GA (FairGA) approach, requires enforcement of certain constraints during the optimization process, in particular a minimum life time for every individual chromosomes, during the optimization process. A preliminary version of a FairGA algorithm is then provided. The disadvantages of the traditional approaches to optimization in terms of their impact outside the space of the optimization objectives have been discussed. The challenges in optimization, especially those related to invisible or immeasurable processes or areas of operation is then considered, and it has been highlighted that an implicit approach is required to control/contain such side effects. The performance of the preliminary version of the FairGA algorithm versus a traditional GA algorithm is then studied on a few standard, benchmark optimization functions. The role and impact of incorporation of the FairGA approach in the operations were then discussed. It has been shown that even for those operations that have recycling of products and circular economy included, the addition of FairGA constraints could greatly reduce the resource extraction and also overhead costs of processing the recycled material.

\section*{Acknowledgment}
The authors thank the NSERC of Canada for their financial support under Grant CRDPJ 424371-11 and under the Canada Research Chair in Sustainable Smart Eco-Cloud (NSERC-950-229052), and also the MITACS of Canada.

\begingroup
\bibliographystyle{IEEEtran} \bibliography{imagep}
\endgroup

\end{document}